\documentstyle[11pt]{article}
\def\fnote#1#2{\begingroup\def\thefootnote{#1}\footnote{#2}
\endgroup}

\begin{document}

\hfill{UTTG-08-00}

\vspace{36pt}

\begin{center}
{\large{\bf { Curvature Dependence of Peaks in the Cosmic Microwave Background 
Distribution
}}}

\vspace{36pt}
Steven Weinberg\fnote{*}{Electronic address:
weinberg@physics.utexas.edu}\\
{\em Theory Group, Department of Physics, University of
Texas\\
Austin, TX, 78712}
\end{center}

\vspace{30pt}

\noindent
{\bf Abstract}

The widely cited formula $\ell_1\simeq 200\,\Omega_0^{-1/2}$ for the multipole 
number of the first Doppler peak is not even a crude approximation in the case 
of greatest current interest, in which  the cosmic mass density is less than the 
vacuum energy density.  For instance, with $\Omega_M$ fixed at 0.3, the position 
of any Doppler peak varies as $\Omega_0^{-1.58}$ near $\Omega_0=1$.
\vfill

\baselineskip=24pt
\pagebreak
\setcounter{page}{1}

The precise measurement$^1$ of the multipole number $\ell_1=197\pm 6$ at the 
first `Doppler' peak has provided an invaluable constraint on cosmological 
parameters.  In a 1994 numerical calculation, Kamionkowski, 
Spergel and Sugiyama$^2$ presented a formula giving $\ell_1$ as a function 
essentially of the curvature alone:
\begin{equation} 
\ell_1\sim \frac{200}{\sqrt{\Omega_0}}\;,
\end{equation} 
where $\Omega_0\equiv \Omega_M+\Omega_\Lambda$, in which $\Omega_M$ and 
$\Omega_\Lambda$ are the  present ratios of the cosmic mass density and the 
vacuum energy (associated, e. g., with a cosmological constant) to the critical 
density. This calculation was done before supernova studies$^3$ 
indicated the likely presence of a relatively large cosmological constant, and 
therefore assumed that $\Omega_\Lambda=0$.  They also explained the $\Omega_0^{-
1/2}$ behavior by noting that $\ell_1$ is approximately inversely proportional 
to the angle subtended at the earth by the horizon at the time of last 
scattering, which was known$^4$ to be proportional to $\Omega_0^{1/2}$ for 
$\Omega_\Lambda=0$.  The same $\Omega_0$-dependence was derived on the same 
grounds by 
Frampton {\em et al.},$^5$ explicitly for the case 
$\Omega_\Lambda=0$.  

Unfortunately, despite the fact that it was derived only for the case 
$\Omega_\Lambda=0$, Eq.~(1) continues to be quoted$^{1,6, 7,8,9}$ as if it were 
generally applicable also when $\Omega_\Lambda$ is appreciable.  As far as I 
know, this formula has not been used by observational groups in analysis of 
their data, but in view of the great current interest in these matters, it seems 
worth warning that {\em in fact,
Eq.~(1) is not valid for parameters in the range suggested by 
supernova observations}, for which $\Omega_\Lambda>\Omega_M$.  Although it 
is true that when $\Omega_0$ is near unity, $\ell_1$ depends less sensitively on 
other parameters than on $\Omega_0$,  the dependence of  $\ell_1$ on  $\Omega_0$ 
bears 
no resemblence whatever to Eq.~(1), except for the case $\Omega_\Lambda\ll 1$.  
Instead, we shall see that the dependence of $\ell_1$ on $\Omega_0$
near $\Omega_0=1$ with $\Omega_M$ fixed at values less than 0.4 is much stronger 
than given by 
Eq.~(1) (for instance, $\ell_1\propto \Omega_M^{-1.58}$ for $\Omega_M=0.3$), and 
it 
depends sensitively on $\Omega_M$.

To calculate the full dependence of $\ell_1$ on $\Omega_0$, $\Omega_M$, 
$\Omega_{\rm baryon}$, $\Omega_{\rm radiation}$, etc. is a complicated task, 
requiring the consideration of the evolution of the acoustic velocity and of the 
ratio of radiation and matter energies, and the consideration of Doppler shifts 
as well as 
temperature fluctuations.  We can avoid all these complications by considering 
the dependence of $\ell_1$ on $\Omega_0$ when only $\Omega_\Lambda$ is allowed 
to vary,  with $\Omega_M$ and all other parameters held fixed.  If it were 
really true (as Eq.~(1) says) that $\ell_1$ depends only on $\Omega_0$,
 then this would be all we need to calculate the full $\Omega_0$-dependence.

The advantage of letting only $\Omega_\Lambda$ vary is that the vacuum energy 
density is negligible compared with 
the densities of matter and radiation at and before  the redshift $z_L\simeq 
1100$ of last scattering, so the only effect of variations in $\Omega_\Lambda$ 
on the multipole number $\ell_n$ of the $n$th Doppler peak is to change the 
paths followed by light rays since the time of last 
scattering.  The angle subtended at the earth by {\em any} feature of the cosmic 
microwave background of proper length $d$ is 
\begin{equation} 
\theta=d/d_A\;,
\end{equation} 
where $d_A$ is the angular diameter distance of the surface of last 
scattering:$^{10}$
\begin{equation}
d_A=\frac{1}{\Omega_k^{1/2}H_0(1+z_L)}\sinh\left[\Omega_k^{1/2}
\int_{1/(1+z_L)}^1\frac{dx}{\sqrt{\Omega_\Lambda x^4+\Omega_k x^2+\Omega_M 
x}}\right]\;,
\end{equation} 
and $\Omega_k$ is a measure of curvature
\begin{equation} 
\Omega_k\equiv 1-\Omega_\Lambda-\Omega_M =1-\Omega_0\;.
\end{equation} 
It follows that the $\Omega_\Lambda$-dependence of $\ell_n$ is given by 
\begin{equation}
\ell_n\propto d_A\;.
\end{equation}
Furthermore, although the relation between the present Hubble constant $H_0$ and 
the proper scales of phenomena at the time of last scattering depends on 
$\Omega_M$ and $\Omega_{\rm radiation}$, it does not depend on $\Omega_\Lambda$.  
(For instance, if we neglect radiation, then the acoustic horizon at the 
redshift of last scattering is $2(1+z_L)^{-3/2}/\sqrt{3\Omega_M}H_0$.)  
Therefore, with $\Omega_M$ fixed,  the dependence of $\ell_n$ on 
$\Omega_\Lambda$ is given by
\begin{equation} 
\ell_n\propto {\cal F}(\Omega_\Lambda)\equiv \frac{1}{\Omega_k^{1/2} 
}\sinh\left[\Omega_k^{1/2}
\int_0^1\frac{dx}{\sqrt{\Omega_\Lambda x^4+\Omega_k x^2+\Omega_M x}}\right]\;,
\end{equation}
with $\Omega_k$ given in terms of $\Omega_\Lambda$ by Eq.~(4).
(The lower limit on the integral has here been set equal to zero because
$z_L>>1$.)  Of course, all the detailed physics of the acoustic oscillations 
responsible for the Doppler peaks is contained in the constant of 
proportionality; all we need to know here is that it does not involve 
$\Omega_\Lambda$.

Now let us consider the variation of the quantity (6) as we make small changes 
in $\Omega_0$ near $\Omega_0=1$ with $\Omega_M$ fixed.  An elementary 
calculation gives
\begin{equation} 
\ell_n\propto \Omega_0^{-\nu}\;,
\end{equation} 
where
\begin{equation} 
\nu\equiv\left(\frac{\partial \ln{\cal F}}{\partial 
\Omega_\Lambda}\right)_{\Omega_\Lambda=1-\Omega_M}=\frac{{\cal I}_1^2}{6}
-\frac{{\cal I}_2}{2{\cal I}_1}\;,
\end{equation} 
with
\begin{equation} 
{\cal I}_1\equiv \int_0^1 \frac{dx}{\left[(1-
\Omega_M)x^4+\Omega_Mx\right]^{1/2}}\;,\qquad
{\cal I}_2\equiv \int_0^1 \frac{(x^2-x^4)\,dx}{\left[(1-
\Omega_M)x^4+\Omega_Mx\right]^{3/2}}\;.
\end{equation}
The table below gives values of these integrals, and of the resulting exponent 
$\nu$ in Eq.~(7).  
 
\begin{center}
\begin{tabular} 
{||l|c|c|c||}  \hline 
$\Omega_M$ & ${\cal I}_1$ & ${\cal I}_2$  & $\nu$ \\ \hline \hline
0.2 & 3.891 & 2.546 & 2.196 \\ \hline
0.3 & 3.305 & 1.601 & 1.578 \\ \hline
0.4 & 2.938 & 1.145 & 1.244 \\ \hline
&&&\\ \hline 
1.0 & 2 & 8/21 & 4/7\\ \hline
\end{tabular}
\end{center}
The only approximation made in deriving these results is that the universe 
becomes transparent suddenly at a redshift
$z_L\gg 1$, and has been dominated since then by non-relativistic matter and 
vacuum energy.  Also, we are neglecting the effect of changing gravitational 
potentials at redshifts $z\ll z_L$, which introduce an additional $\Lambda$-
dependence$^{11}$ that is quite small at the wavelengths of the Doppler peaks.  
Otherwise, these results are exact.

The behavior $\ell_1\propto \Omega_0^{-4/7}$ near $\Omega_0=1$ for $\Omega_M$ 
fixed at unity is  close to the behaviour $\ell_1\propto \Omega_0^{-1/2}$ near 
$\Omega_0=1$ found$^{2,5}$ for $\Omega_\Lambda$ fixed at zero, confirming that 
$\ell_1$ is approximately a function of $\Omega_0$ alone for 
$\Omega_\Lambda=0$ and $\Omega_M$ near  unity.  The fact that $\nu$ depends 
strongly on $\Omega_M$ for 
smaller values of $\Omega_M$ shows that for observationally favored parameters 
$\ell_1$ is {\em not} approximately a function of $\Omega_0$ alone.  Indeed, 
there 
is no physical reason why $\ell_1$ {\em should} be even approximately a 
function of $\Omega_0$ alone.
For fixed values of $\Omega_M$ less than 0.4 
the $\ell_n$ fall off with increasing $\Omega_0$ much more rapidly than 
would be expected from Eq.~(1), so the measurement of the positions of the 
Doppler 
peaks provides a more stringent constraint on $\Omega_0$ than would be the case 
if Eq.~(1) were correct.

\begin{center}
{\bf Acknowledgements}
\end{center}

I am grateful for conversations with M. Kamionkowski, M. Roos, M. Turner, and M. 
White. This research was supported in part by the Robert A. Welch Foundation and 
NSF Grant PHY-9511632.

\begin{center}
{\bf References}
\end{center}
\begin{enumerate}
\item P. de Bernardis {\em et al.}, {\em Nature} {\bf 404}, 955 (2000).
\item M Kamionkowski, D. N. Spergel, and N. Sugiyama, {\em Ap. J.} {\bf 426}, 
L57 (1994).
\item S. Perlmutter {\em et al.}, astro-ph/9812133, 9812473; B. P. Schmidt {\em 
et al.}, {\em Ap. J.} {\bf 507}, 46 (1998); A. G. Riess {\em et al.}, 
astro-ph/9805200.
\item See, e. g., S. Weinberg, {\em Gravitation and Cosmology} (Wiley, New York, 
1972), Eq.~(15.5.39); E. W. Kolb and M. S. Turner, {\em The Early Universe} 
(Addison-Wesley, Redwood City, CA, 1990), p. 505.
\item P. Frampton , Y. J. Ng, and R. Rohm, {\em Mod. Phys. Lett.} 
{\bf A13}, 2541 (1998).  There are aspects of this paper with which I disagree, 
but they are not relevant to the present work. 
\item N. A. Bahcall, J. P. Ostriker, S. Perlmutter, and P. J. Steinhardt, {\em 
Science} {\bf 28}, 1481 (1999).
\item M. S. Turner, in {\em Cosmo-98: Second International Workshop on Particle 
Physics and the Early Universe}, AIP Conference Proceedings {\bf 478} (American 
Institute of Physics, Woodbury, NY 1999), ed. by D. O. Caldwell, p. 113.
 \item M. Roos and S. M. Harun-or-Rashid, astro-ph/0005541 (2000).
\item Bahcall {\em et al.}$^6$ cited reference 2, while Turner$^7$ cited no 
reference 
for Eq.~(1).  De Bernardis {\em et al.}$^1$ cited no references for Eq.~(1), but 
relied on references 2, 5, and 6.  Roos and Harun-or-Rashid$^8$ also cited no 
references, but took this formula from reference 1. 
\item This formula is given, e. g., in reference 5.
\item M. White, D. Scott, and J. Silk, {\em Ann. Rev. Astron. Astrophys.} {\bf 
32}, 319 (1994): Appendix B; W. Hu and M. White, {\em Astron. Astrophys.} {\bf 
315}, 33 (1996).
\end{enumerate}
\end{document}